\documentclass[paper]{JHEP3} 

\JHEPspecialurl{http://jhep.sissa.it/JOURNAL/JHEP3.tar.gz}

\usepackage{epsfig,multicol}

\newcommand\fverb{\setbox\pippobox=\hbox\bgroup\verb}
\newcommand\fverbdo{\egroup\medskip\noindent%
			\fbox{\unhbox\pippobox}\ }
\newcommand\fverbit{\egroup\item[\fbox{\unhbox\pippobox}]}
\newbox\pippobox
\newcommand{\beq}{\begin{equation}}
\newcommand{\eeq}{\end{equation}}

\title{Confining Strings at High Temperature}

\author{$^a$M.C. Diamantini\thanks{Supported by a Swiss National Science Foundation
Fellowship. On leave of absence from I.N.F.N. and University of Perugia, Italy.}~
		and $^b$C.A.Trugenberger\\
	$^a$$^b$Theory Division, CERN, CH-1211 Geneva 23, Switzerland\\
	$^b$InfoCodex SA, av. Louis-Casa\"\i\ 18,
	CH-1209 Geneva, Switzerland\\
	E-mail: \email{cristina.diamantinitrugenberger@cern.ch},
	 \email{ca.trugenberger@InfoCodex.com}}
\received{} 		
\revised{}
\accepted{}		

\preprint{}	 \preprint{CERN-TH/2002-055\\March 2002}

\abstract{We show that  the  high-temperature behaviour of the recently proposed
confining strings reproduces exactly the correct large-$N$ QCD result, for a
large class of truncations of the long-range interaction between surface elements.}


\begin{document} 


\section{Introduction}
Coloured states are not observable as free particles: at large distances the static
potential between a quark and an antiquark grows linearly with the distance.
Attempting to explain this quark confinement by means of a
non-critical string theory dual to QCD remains an open and very important problem
 \cite{polc1}. Indeed 
finding the  theory dual to  QCD would imply  having perturbative control over both
the asymptotically free and the confinement regimes of the theory.
Despite many efforts, however,  the formulation of a consistent string theory
description of confinement in 4D non-Abelian gauge
theories remains an open issue.
The Nambu--Goto term can be consistently quantized only in $D = 26$ or $D \leq 1$ and leads
to strings with a crumpled world-sheet, inappropriate to describe the expected smooth strings
dual to QCD \cite{poly1} .
In order to take into account the bending rigidity of flux tubes, and to cure the
problems of the fundamental Nambu--Goto action, a term proportional to the extrinsic curvature
of the world-sheet was added to it \cite{poly}. This term, however, is infrared-irrelevant and the
rigid string thus shares the  geometrical problems of the Nambu--Goto action
\cite{poly}.

A new  action for confining strings has recently been proposed in
\cite{pima}.
The confining string action is based on an induced string
action explicitly derivable for compact QED \cite{pimo} and for
Abelian-projected SU(2) \cite{anto}, and possesses, in its world-sheet formulation, a non-local
action with a negative stiffness \cite{pimo,klei} that can be expressed as a
derivative expansion of a long-range interaction between surface elements.
In order for the geometrical properties of these strings to be analytically
studied, the expansion must be truncated: since the stiffness is negative, a
stable truncation must, at least, include a sixth-order term in the derivatives
\cite{cris1}.
This string action has many features that make it a good candidate to describe QCD flux tubes,
at least in the large-$D$ approximation. In fact this model has an infrared fixed point at zero
stiffness \cite{cris1,cris2}, corresponding to a
tensionless smooth string whose world-sheet has Hausdorff dimension 2, exactly
the desired properties to describe QCD flux tubes.
The long-range orientational order in 
this model is due to an antiferromagnetic interaction between normals to the surface \cite{cris3}, a mechanism
confirmed by numerical simulations \cite{chern}.
Moreover, it was shown in \cite{cris2} that this infared fixed point does not depend on the
truncation and is present for all ghost- and tachyon-free truncations, and that
the effective theory describing the infrared behaviour is a conformal field
theory with central charge $c = 1$.

There is, however another characteristic that any string model describing 
confinement in QCD must possess, namely the correct high-temperature behaviour.
In fact, as shown in \cite{polc3},
the deconfining transition in QCD is due to the condensation of Wilson lines, and the
partition function of QCD flux tubes can be continued above the deconfining
transition; this high-temperature continuation can be evaluated
perturbatively. So, any string theory that is equivalent to QCD {\it must} reproduce
this behaviour. 
As  shown in \cite{cris4}, contrary to previous string models, 
the high-temperature behaviour of 
confining strings reproduces exactly the correct large-$N$ QCD result, a {\it necessary}
condition for any string model of confinement.

In \cite{cris4}, the correct high-temperature behaviour of the confining string action
was derived for a specific truncation. In this paper we will show that 
this property is largely independent
of the specific truncation we choose, but  holds true for 
all ghost- and tachyon-free truncations.

\section{Definition of the Model}

The model proposed in \cite{cris1} is
\beq
S = \int d^2 \xi \sqrt{g}  g^{ab} {\cal D}_a x_\mu \left( t - s{\cal D}^2 +
 {1 \over M^2} {\cal D}^4 \right){\cal D}_b x_\mu  \ , \label{model}
 \eeq
where ${\cal D}_a$ are covariant derivatives with respect to the induced metric
$g_{ab} = \partial_a x_\mu \partial_b x_\mu$ on the surface ${\bf x}(\xi_0,
\xi_1)$.
The first term in the bracket provides a bare surface tension $2t$, while the
second accounts for the rigidity, with a stiffness parameter $s$ that can be positive or
negative. The third term contains
the square of the gradient of the extrinsic curvature matrices, and  it suppresses the
formation of spikes on the world-sheet. The new mass scale $M$ generates, 
in the large-$D$ approximation, 
string tension proportional to $M^2$, which takes control of the
fluctuations where the orientational correlation dies off.

We now want to generalize this model to an arbitrary truncation of the long-range interaction
in (\ref{model}).
Following \cite{cris2} we will consider an action of the form:
\begin{eqnarray}
S|_n &&= \int d^2 \xi \sqrt{g}  g^{ab}{\cal D}_a x_\mu  V_n ({\cal D}^2){\cal D}_b x_\mu \ ,
\nonumber \\
V_n ({\cal D}^2) &&= t \Lambda^2 + \sum_{k=1}^{2n} {c_k \over \Lambda^{2k-2} } ({\cal
D}^2)^k\ ,
\label{nmodel}
\end{eqnarray}
where $\Lambda$ represents the fundamental mass scale in the model, to be identified with the QCD
mass scale; $c_k$ are positive numbers, which means that
a stable truncation must end with an even $ k = 2n$.
The model studied in \cite{cris4} corresponds to $n=1$.

The parameters $c_k$ are free: the only condition we impose on them 
 is the absence of both tachyons and ghosts in the theory.
This requires that the Fourier  transform $V_n\left( p^2 \right)$ have
no zeros on the real $p^2$-axis. The polynomial $V_n \left( p^2 \right)$ thus 
has $n$ pairs of complex-conjugate zeros in the complex $p^2$-plane \cite{cris2}.
To simplify the computation we will also set all coefficients with odd $k$
to zero, $c_{2m+1} =0$ for $0\le m \le n-1$. This, however, is
no drastic restriction since, as was shown in \cite{cris2}, this is
their value at the infrared-stable fixed point anyhow. 

We will study this model in the large-$D$ approximation. For this purpose \cite{david}
we need to introduce  a Lagrange multiplier 
$ \lambda^{ab}$  that forces the induced metric  
$\partial_a x_\mu \partial_b x_\mu$ to be equal to the intrinsic metric $g_{ab}$, extending the
action (\ref{model}) to:
\beq
S|_n\rightarrow S|_n + \int d^2 \xi \sqrt{g} \left[\Lambda^2\lambda^{ab} (\partial_a x_\mu \partial_b x_\mu - g_{ab}
)  \right] \ . \label{cmodel} 
\eeq
Note that we define here the Lagrange multiplier $\lambda^{ab}$ as a dimensionless quantity.
We parametrize the world-sheet in a Gauss map by $x_\mu (\xi) = (\xi_0, \xi_1,
\phi^i(\xi)),\ i = 2,...,D-2$. The value of the periodic coordinate $\xi_0$ is
$-\beta/2 \leq \xi_0 \leq \beta/2$, with $\beta = 1/T$ and $T$  the temperature.
The value of $\xi_1$  is
$-R/2 \leq \xi_1 \leq R/2$; $\phi^i(\xi)$ describe the $D-2$ transverse
fluctuations. We look for a saddle-point solution with a diagonal metric $g_{ab}
= {\rm diag}\ (\rho_0, \rho_1)$, and a Lagrange multiplier of the form
$\lambda^{ab} = {\rm diag}\ (\lambda_0/\rho_0, \lambda_1/\rho_1)$.
The action then becomes:
\begin{eqnarray}
S|_n &&= S|_{n,0} + S|_{n,1}\nonumber \\
S|_{n,0} &&=  A_{\rm ext}\ \Lambda^2 \sqrt{\rho_0 \rho_1} \left[ \right. t \left( {\rho_0 + \rho_1
\over \rho_0 \rho_1} \right) + \lambda_0 \left( { 1 -\rho_0 \over \rho_0}
 \right) 
+ \lambda_1 \left( { 1 -\rho_1 \over \rho_1} \right) \left. \right] \nonumber \ ,\\
S|_{n,1} &&= \int d^2 \xi \sqrt{g} \left[ g^{ab}\partial_a \phi^i  V_n ({\cal D}^2) 
\partial_b \phi^i + \Lambda^2 \lambda^{ab} \partial_a \phi^i \partial_b \phi^i \right]\ ,
\end{eqnarray}
where $\beta R = A_{\rm ext}$ is the extrinsic, projected area in coordinate
space, and $S|_{n,0}$ is the tree-level contribution.
 Integrating over the transverse fluctuations in the one-loop term $S|_{n,1}$, we obtain, in the limit $R \to
\infty$:
\beq
S|_{n,1} = {D - 2\over 2} R \sqrt{\rho_1} \sum_{l = - \infty}^{+ \infty} \int {d
p_1\over 2 \pi} \ln\left[ 
 (p_1^2 \lambda_1 + \omega_l^2 \lambda_0) \Lambda^2 + p^2 V_n(p^2)  \right] \ ,
\eeq
where $p^2 = p_1^2 + \omega_l^2$, and  $\omega_l = {2 \pi \over \beta \sqrt{\rho_0} } l$.

\section{High-Temperature Behaviour}
At high temperatures, satisfying
\beq 
{c_{2n}\over \Lambda^{4n-2}} {1\over \beta^{4n}} \gg \Lambda^2 t + \sum_{ k = 1 }^{2n -1} {c_{k} \over \Lambda^{2k-2}}
{1\over \beta^{2k}}\ ,
\label{valid}
\eeq
with $k$ even, the highest-order term in the derivatives dominates in the one-loop term $S|_{n,1}$
when $ n \neq 0$. 
Using analytic regularization and analytic continuation of the formula $\sum_{n =
1}^\infty n^{-z} = \zeta(z)$,
 for the Riemann zeta function, with $\zeta(-1) = -
1/12$, we obtain for the $ n \neq 0$ contribution:
\begin{eqnarray}
&&{D - 2\over 2} R \sqrt{\rho_1} \sum_{l = - \infty}^{+ \infty} \int {d
p_1\over 2 \pi}\ {\rm ln}\ {c_{2n}\over \Lambda^{4n-2}}
\left( \omega_l^2 + p_1^2 \right)^{2n +1}\nonumber \\
&&= {D - 2\over 2}\sqrt{\rho_1 \over \rho_0} (2n + 1) 4 \pi {R\over \beta} 
\sum_{l = 1}^{+ \infty}\sqrt{l^2}
=  -{D - 2\over 2}\sqrt{\rho_1 \over \rho_0}{(2n + 1) \pi \over3}{ R\over \beta}\ .
\end{eqnarray}

The calculation of the $n=0$ contribution is a little more involved.
For $n=0$ we can rewrite $S|_{n,1}$ as:
$$
S|_{n,1} = {D - 2\over 2} R \sqrt{\rho_1}  \int {d
p_1\over 2 \pi} \ln \left( p_1^2 \bar V_n(p_1^2)\right) \ ,
$$
with
\beq
\bar V_n(p_1^2) = 
\left( \Lambda^2 (t + \lambda_1) +  \sum_{k =1}^{2n} 
{c_{k} \over \Lambda^{2k-2}} p_1^{2k}\right)   \ ,\label{newpot}
\eeq
and $k$ even.
With the simplification we have introduced for the $c_k$, namely that 
all coefficients with odd $k$ are zero,
and the requirement that the model is ghost- and tachyon-free,
 all pairs of complex-conjugate zeros of $\bar V_n\left( p_1^2
\right)$ lie on the imaginary axis and we can represent $\bar V_n
\left( p_1^2 \right)$ as \cite{cris2}
\beq
{\Lambda^{4n-2}\over c_{2n}} \ \bar V_n\left( p_1^2
\right) = \prod_{k=1}^n \left( p_1^4 + \alpha_k^2 \Lambda^4 \right) \ ,
\label{poten}
\eeq
with purely numerical coefficients $\alpha_k$.
The $n=0$ contribution then becomes:
\begin{eqnarray}
S|_{n,1} &&= {D - 2\over 2} R \sqrt{\rho_1} \sum_{k = 1}^{n} \int {d
p_1\over 2 \pi} \ln\left( p_1^4   + \alpha_k^2 \Lambda^4  \right) \nonumber \\
&&= {D - 2\over 2} R \sqrt{\rho_1} \sum_{k = 1}^{n} \int {d
p_1\over 2 \pi} 2 {\rm Re} \ln\left( p_1^4   + i \alpha_k \Lambda^2 \right) = 
{D - 2\over 2} R \sqrt{\rho_1} \sum_{k = 1}^{n} \Lambda \sqrt{2 \alpha_k} \ ,
\end{eqnarray} 
and we obtain a total action of the form:
\beq
S_n = S|_{n,0} + {D - 2\over 2} R \sqrt{\rho_1} \left[\sum_{k = 1}^{n} \Lambda 
\sqrt{2 \alpha_k} -
{(2n + 1) \pi \over 3 \sqrt{\rho_0}}{ 1\over \beta}\right] \ .
\label{finac}
\eeq

Note that the $p_1$-independent term in (\ref{poten})  must satisfy:
\beq
\prod_{k=1}^n \alpha_k^2 \Lambda^4 = {\Lambda^{4n}\over c_{2n}}(t + \lambda_1)\ .
\label{gamcon}
\eeq
Following \cite{cris2}, we assume that all $\alpha_k$ are equal  so that (\ref{gamcon}) implies:
\beq
\alpha^2_k = ( t + \lambda_1)^{1/n} \alpha^2\ ,\ \ \ \alpha = \left({1\over c_{2n}}\right)^{1/2n}
\label{coeff}
\eeq
With this assumption we obtain the four large-$D$ gap equations:
\begin{eqnarray}
&&{ 1 -\rho_0 \over \rho_0} =  0 \ , \label{gap1}\\
&&{1 \over \rho_1} = 1 - {D - 2\over 2} { 1 \over 4 \beta \Lambda} 
\sqrt{2 \alpha} (\lambda_1 + t )^{1/4n -1} \ , \label{gap2} \\
&&\left[ {1\over 2}(t - \lambda_1) + {1 \over 2\rho_1}(\lambda_1 + t ) - t -
\lambda_0 \right] + {D - 2\over 2} { (2n +1)\pi \over 6 \beta^2 \Lambda^2} = 0 \ ,\label{gap3} \\
&&(t - \lambda_1) - {1 \over \rho_1}(\lambda_1 + t ) + 
{D - 2\over 2} {1 \over  \beta \Lambda} \left[\sqrt{2 \alpha}\ n \left( \lambda_1 
+ t \right)^{1/4n} - {\pi (2n+1) \over 3 \beta \Lambda } \right] = 0 \label{gap4} \ .
\end{eqnarray}
Inserting (\ref{gap4}) and (\ref{gap1}) into (\ref{finac}) and using $\rho_0 =1$ from
(\ref{gap1}) we obtain a simplified form of the
effective action:
\beq
S^{\rm eff} = A_{\rm ext}\ \Lambda^2 {\cal T} \sqrt{1\over \rho_1}
\label{effac} \ ,
\eeq
with ${\cal T} =  2 (\lambda_1 + t )$  representing  the physical string
tension.

Without loss of generality we now set 
\beq
\sqrt{2 \alpha} = \gamma (\lambda_1 + t )^{-1/4n +1/2} \ . 
\label{ans}
\eeq
Inserting (\ref{gap3}) into (\ref{gap4}) we then obtain an equation for $(\lambda_1 + t )$
alone:
\beq
(\lambda_1 + t ) -  {D - 2\over 2} {4 n +1 \over 8 \beta \Lambda}\gamma
\left( \lambda_1 + t \right)^{1/2} +\ {D - 2\over 2}{2 n +1 \over  3} 
{ \pi \over   2\beta^2\Lambda^2 } - t = 0 \ .
\label{quart}
\eeq
This equation has two solutions:
\begin{eqnarray}
(\lambda_1 + t )^{1/2} &&= {D- 2 \over 2}{4 n +1 \over  16 \beta \Lambda} \times
\nonumber \\
&&\times \left[ 1
\pm \sqrt{ 1 - {2 n +1 \over  3} {128 \pi \over (4 n +1)^2 \gamma^2{D- 2 \over 2}} +
 {256\ t\ \beta^2\  \Lambda^2 \over (4 n +1)^2 \gamma^2\left({D- 2 \over 2}\right)^2}} \right] \ .
\label{solut}
\end{eqnarray}
When 
\beq 
\gamma >  \left({D- 2 \over 2}\right)^{-1/2} {8 \over  (4 n +1)} \sqrt{2(2 n +1) \pi
\over  3} \ ,
\label{posdis}
\eeq
the solutions (\ref{solut}) are both real, independently of $\beta$.
As in \cite{cris4}, for the solution with the plus sign, $1/\rho_1$ is always positive
(note that  $n \geq 1$):
\beq
{1\over \rho_1} = 1 - {4 \over (4n +1)\left[ 1 + \sqrt{ 1 - {2 n +1 \over  3} 
{128 \pi \over (4 n +1)^2 \gamma^2{D- 2 \over 2}} +
{256\ t\ \beta^2\ \Lambda^2  \over (4 n +1)^2
 \gamma^2\left({D- 2 \over 2}\right)^2}}\right]}\ .
\label{posrho}
\eeq
The square of the free energy $F^2(\beta) \equiv {S^2_{\rm Eff} \over R^2}$ is thus positive,
since $(\lambda_1 + t )$ is real:
\begin{eqnarray}
&&F^2(\beta)  = {1\over \beta^2}\ \left({D- 2 \over 2}
{(4 n +1)  \over  16}\right)^4 \nonumber \\
&&\left[ 1
 -\sqrt{ 1 - {2 n +1 \over  3} {128 \pi \over (4 n +1)^2 \gamma^2{D- 2 \over 2}} +
 {256\ t\ \beta^2\  \Lambda^2 \over (4 n +1)^2 \gamma^2\left({D- 2 \over 2}\right)^2}} \right]^4\
 \times \nonumber \\
\times &&\left[ 1 - {4 \over (4n +1)\left[ 1 + \sqrt{ 1 - {2 n +1 \over  3} {128 \pi \over (4 n +1)^2 \gamma^2{D- 2 \over 2}} +
{256\ t\ \beta^2\  \Lambda^2 \over (4 n +1)^2 
\gamma^2\left({D- 2 \over 2}\right)^2}}\right]} \right] \ . 
\label{freew}
\end{eqnarray}
As for the rigid string \cite{polc4}, the high-temperature behaviour is the same as in QCD, 
but the sign is wrong.
The crucial difference with respect to the rigid string case is that (\ref{freew}) is real, while the
squared free energy for the rigid string is imaginary, signalling an 
instability in the model \cite{polc4}.
If we now look at the behaviour of $\rho_1$ at low temperatures, below the deconfining transition
\cite{cris2}, we see that $1/\rho_1$ is positive. The deconfining transition is indeed 
determined by the
vanishing of $1/\rho_1$ at $\beta = \beta_{\rm dec}$. In the case of
(\ref{posrho}), this means that   $1/\rho_1$ is positive below the Hagedorn transition, touches
zero at $\beta_{\rm dec}$ and remains positive above it. Exactly the same will happen also for 
$F^2$, which is positive below $\beta_{\rm dec}$ , touches
zero at $\beta_{\rm dec}$ and remains positive above it.
This solution thus describes an unphysical ``mirror'' of the low-temperature behaviour of the
confining string, without a real deconfining Hagedorn transition. For this reason we discard it.

We will now concentrate on the solution of (\ref{solut}) with the minus sign.
In this case we have
\beq
{1\over \rho_1} = 1 - {4 \over (4n +1) - (4n +1)\sqrt{ 1 - {2 n +1 \over  3} {128 \pi \over (4 n +1)^2 \gamma^2{D- 2 \over 2}} +
{256\ t\ \beta^2\ \Lambda^2 \over (4 n +1)^2 \gamma^2\left({D- 2 \over 2}\right)^2}}}\ .
\label{negrho}
\eeq
If 
\beq  
\gamma >  \left({D- 2 \over 2}\right)^{-1/2} 4 \sqrt{2n +1 \over 4 n -1} \sqrt{ \pi
\over  3}\ ,
\label{condalp}
\eeq
$1/\rho_1$ is negative independently, of $\beta$. Note that, when (\ref{condalp}) is satisfied, then 
also  (\ref{posdis}) is automatically satisfied. We will restrict to this range of values
for $\gamma$.
In this case, 
as in \cite{cris4}, the physical string tension is real and proportional to $1/\beta^2$
and $1/\rho_1$ is negative, giving: 
\begin{eqnarray}
&&F^2(\beta)  = -{1\over \beta^2}\ \left({D- 2 \over 2}
{(4 n +1)  \over  16}\right)^4 \nonumber \\
&&\left[ 1
 -\sqrt{ 1 - {2 n +1 \over  3} {128 \pi \over (4 n +1)^2 \gamma^2{D- 2 \over 2}} +
 {256\ t\ \beta^2\  \Lambda^2 \over (4 n +1)^2 \gamma^2\left({D- 2 \over 2}\right)^2}} \right]^4\
 \times \nonumber \\
\times &&\left[{4 \over (4n +1)\left[ 1 - \sqrt{ 1 - {2 n +1 \over  3} {128 \pi \over (4 n +1)^2 \gamma^2{D- 2 \over 2}} +
{256\ t\ \beta^2\  \Lambda^2 \over (4 n +1)^2 
\gamma^2\left({D- 2 \over 2}\right)^2}} - 1 \right]} \right] \ . 
\label{freec}
\end{eqnarray}
In the range defined by (\ref{condalp}), this is {\it negative}. Note, moreover, that for
$n=1$ we exactly reproduce the result found in \cite{cris4}.
Since the sign of $\lambda_1$ does not change at high temperatures, the field $x_\mu$ is not
unstable. In the rigid string case \cite{polc4}, instead, the change of
sign of $\lambda_1$ gives rise to a world-sheet instability.

Let us now compare with the large-$N$ QCD result \cite{polc3}: 
\beq
F^2(\beta)_{\rm QCD} = - {2 g^2(\beta) N \over \pi^2 \beta^2}\ ,\label{qcd}
\eeq
where $g^2(\beta)$ is the QCD coupling constant. To this end we first simplify our result 
by considering 
$$
\gamma \gg  \left({D- 2 \over 2}\right)^{-1/2} {8 \over (4 n +1)} \sqrt{2(2 n +1) \pi
\over  3}\ .
$$
Moreover we ask that
\beq
t \beta^2 \ll { D-2 \over 2} \ ,
\label{cont}
\eeq
so that we can ignore the last term in the square root in (\ref{freec}).
Note that this condition is compatible with the high-temperature approximation
(\ref{valid}).
In this case (\ref{freec}) reduces to 
\beq
F^2(\beta) = - {1\over \beta^2 } {4 \pi^3 \over 125}{(2n +1)^3 \over (4n +1)^2} 
{D-2\over \gamma^2}\ .
\label{bingo}
\eeq
This corresponds {\it exactly} to the QCD result (\ref{qcd}), with the identifications
\begin{eqnarray}
g^2 &&\propto {1\over  \gamma^2} \ , \nonumber \\
N && \propto D - 2 \ . \nonumber
\end{eqnarray}
The weak $\beta$-dependence of the QCD coupling $g^2(\beta)$ can be accommodated in the parameter
$\gamma $. As one would expect for an asymptotic free theory \cite{wilczek}, our
high-temperature result is valid at large values of $\gamma$, 
i.e. small values of $g^2$. As first noted in \cite{cris4} there is an interesting relation
between the order of the gauge group and the number of transverse space-time dimensions.

Together with the absence of crumpling, this result is a very strong indication that
non-local interactions between surface elements are crucial to describe the world-sheet of
QCD flux tubes. Investigations of the fundamental string model in which these interactions
are mediated by an antisymmetric tensor field are in progress.

\bigskip


\end{document}